\documentstyle[12pt,epsf]{article}
\def\be{\begin{equation}}
\def\ee{\end{equation}}
\def\bea{\begin{eqnarray}}
\def\eea{\end{eqnarray}}

\def\br{\begin{thebibliography}{abc}}
\def\er{\end{thebibliography}}

\def\bar#1{\overline{#1}}

\def\a{\alpha}
\def\b{\beta}

\newcommand{\ZZ}{ Z \!\!\! Z}
\newcommand{\comm}{{\rm l}\!\!\!{\rm C}}
\def\real{\relax{\rm I\kern-.18em R}}
\def\com{\relax\,\hbox{$\inbar\kern-.3em{\rm C}$}}

\def\br{{\bf R}}

\def\inbar{\,\vrule height1.5ex width.4pt depth0pt}
\def\IC{\relax\,\hbox{$\inbar\kern-.3em{\rm C}$}}
\def\ID{\relax{\rm I\kern-.18em D}}
\def\IF{\relax{\rm I\kern-.18em F}}
\def\IH{\relax{\rm I\kern-.18em H}}
\def\II{\relax{\rm I\kern-.17em I}}
\def\I1{\relax{\rm 1\kern-.28em l}}
\def\IN{\relax{\rm I\kern-.18em N}}
\def\IP{\relax{\rm I\kern-.18em P}}
\def\IQ{\relax\,\hbox{$\inbar\kern-.3em{\rm Q}$}}
\def\IZ{\relax\,\hbox{$\inbar\kern-.3em{\rm Z}$}}
\def\R{\relax{\rm I\kern-.18em R}}
\font\cmss=cmss10 \font\cmsss=cmss10 at 7pt
\def\Z{\relax\ifmmode\mathchoice
{\hbox{\cmss Z\kern-.4em Z}}{\hbox{\cmss Z\kern-.4em Z}}
{\lower.9pt\hbox{\cmsss Z\kern-.4em Z}}
{\lower1.2pt\hbox{\cmsss Z\kern-.4em Z}}\else{\cmss Z\kern-.4em
Z}\fi}
\def\ca{{\cal A}}
\def\cb{{\cal B}}
\def\cc{{\cal C}}

\def\ch{{\cal H}}

\def\cn{{\cal N}}
\def\co{{\cal O}}

\def\bar#1{\overline{#1}}

\def\Hat#1{\rlap{\kern.10em$\widehat{\phantom G}$}#1}
\def\HAt#1{\rlap{\kern.05em$\widehat{\phantom G}$}#1}

\def\czp#1{\rlap{\kern.1em$\widehat{\phantom{G\vrule height.8em}}$}#1{}}
\def\Czp#1{\rlap{\kern.05em$\widehat{\phantom{G\vrule height.8em}}$}#1{}}

\newcommand{\sect}[1]{\setcounter{equation}{0}\section{#1}}

\def\sxn#1{\bigskip\medskip \sect{#1} \smallskip
                                                 }

\begin{document}

\thispagestyle{empty}
\setcounter{page}{0}


\begin{flushright}
SU-4240-657\\
February 1997
\end{flushright}

\vglue 0.6cm

\centerline {{\Large{\bf Bringing Up a Quantum Baby}}\footnote{Talk
presented
at the Workshop on Frontiers in Field Theory, Quantum Gravity and
String Theory, Puri, December 12-21, 1996.}}

\vglue 0.6cm
                        
\centerline {\large  A.P. Balachandran}
\vglue 0.5cm
{\centerline{\it {Department of Physics, Syracuse University,}}
\centerline{\it {Syracuse, NY 13244-1130, USA}}

\vglue 1cm

\centerline {\bf Abstract}
\vglue 0.6cm

     Any two infinite-dimensional (separable) Hilbert spaces are
unitarily
isomorphic.  The sets of all their self-adjoint operators are also
therefore unitarily
equivalent.  Thus if all self-adjoint operators can be observed, and if
there is no further major axiom in quantum physics than those formulated
for example in Dirac's `Quantum Mechanics', then a quantum physicist
would not be able to tell a  torus from a hole in the ground.  We argue
that there are indeed such axioms involving vectors in the domain of the
Hamiltonian:  The   ``probability densities'' (hermitean forms)
$\psi^\dagger \chi$ for $\psi,\chi$ in
this
domain
generate an algebra from which the classical configuration space with its
topology (and with further refinements of the axiom, its $C^K-$ and
$C^\infty-$
structures) can be reconstructed using Gel'fand - Naimark theory.
Classical topology is an attribute of only certain quantum states for
these axioms, the configuration space emergent from quantum physics
getting progressively less differentiable with increasingly higher
excitations of energy and eventually altogether ceasing to exist.  After
formulating these axioms, we apply them to show the possibility of
topology change and
to discuss quantized fuzzy topologies.  Fundamental issues concerning the
role of time in quantum physics are also addressed.

\newpage

\vglue 0.6cm                     
\sxn{{\bf Introduction}}
\vglue 0.5cm
	
	This note will report on certain ongoing research with several
colleagues concerning the nature of space and time in quantum physics.
Some of our ideas have already been published elsewhere
\cite{bbellst,bbms,bc}. Our
work touches both on issues of relevance to quantum gravity such as the
meaning of ``quantized topology'' and the possibility of topology change,
and on topics of significance for foundations of quantum physics.  I
think that we have  progressively approached a measure of precision in
the formulation of relatively inarticulated questions, but our responses
are still  tentative and lacking in physical and mathematical
completeness and rigor.

\vglue 0.6cm
\sxn{ {\bf   The Problem as a Parable}}
\vglue 0.5cm

The problem to be addressed here is best introduced as a little story
about a quantum baby.  The story will set the framework for the rest of
the talk.  Its proper enjoyment calls for a willing suspension of
disbelief for the moment.

All babies are naturally quantum, so my adjective for the baby can be
objected to as redundant and provocative, but it invites attention to a
nature of infants of central interest to us, so let us leave it there.

\vglue 0.6cm
\noindent {\bf Parable of the Quantum Baby}
\vglue 0.5cm

Entertain the conjecture of a time, long long ago, when there lived a
quantum baby of cheerful semblance and sweet majesty.  It was brought up
by its doting parents on a nourishing diet of self-adjoint operators on a
Hilbert space.  All it could experience as it grew up were their mean
values in quantum states.  It  did not have a clue when it was little
that there is our classical world with its topology, dimension and
metric.  It could not then tell a torus from a hole in the  ground.

Yet the baby learned all that as it grew up.  

And the wise philosopher is
struck with wonder:  How did the baby manage this amazing task?  
 
For the problem is this: Even in a quantum theory emergent  from a smooth
classical configuration space $Q^\infty$, there is no need for a wave
function $\psi$, or a probability density $\psi^*\psi$, to be continuous
on
$Q^\infty$.  It is enough that the integral $\int \omega \psi^*\psi$ over
$Q^\infty$ for an appropriate volume form $\omega$ is finite.
Probability
interpretation  requires no more.

But if the baby can observe all self-adjoint operations with equal ease,
and thereby prepare all sorts of discontinuous quantum states, how then
does it ever learn of $Q^\infty$, its topology and its differential
attributes?

This then is our central question.  All that follows is charged with its
emotional content, and comes from trying to find its answer.

\vglue 0.6cm
\sxn{{\bf  Another Formulation}}
\vglue 0.5cm

We can explain the baby problem in yet another way.

In classical physics, we generally deal with dynamics on a
$(C^{\infty}-)$
manifold, the classical configuration space $Q^\infty$. Observables are
(suitably) smooth functions in the cotangent bundle $T^*Q^\infty$ while
states are similarly smooth probability distributions on that space.
Time evolution preserves these smooth structures.  There is no particular
need to go beyond them and abandon the emphasis on smoothness.  In this
sense, classical physics incorporates the ideas of underlying manifolds,
it knows about them all along.  They are generally {\em a priori} and
irreducible concepts
for classical physics, with no model for their emergence and immune to
analysis,  but all the same
essential in the formulation of its physical laws. 

In quantum physics
instead,
time evolution is  specified by a unitary operator $U(t)$, continuous
in
$t$, on a (separable \cite{rs}) Hilbert space $\ch$.  [We will deal only
with
separable Hilbert spaces.]  The latter is generally infinite-dimensional.

But unlike $Q^\infty$, all infinite-dimensional Hilbert spaces are
isomorphic, in fact unitary so.  If $|n>^{(i)}(n\in {\IN})$ gives the
orthonormal basis for the Hilbert space $\ch^{(i)} (i=1,2)$, we can
achieve this equivalence by setting $|n>^{(2)}=V|n>^{(1)}$.  That being
so, any operator $A^{(1)}$ on $\ch^{(1)}$ has a corresponding operator 
$A^{(2)}= VA^{(1)}V^{-1}$ on $\ch^{2)}$. 

How then does a quantum baby tell a torus from a hole in the ground?

{\em Without further structure in quantum physics besides those to
be found in standard text books, this task is in fact entirely beyond the
baby.}

In conventional quantum physics, we generally start from smooth functions
(or smooth sections of hermitean vector bundles) on $Q^\infty$ and
complete them into a Hilbert space $\ch$ using a suitable scalar product.
In this way, we somehow incorporate knowledge about $Q^\infty$ right at
the start.

But this approach requires realizing $\ch$ in a particular way, as square
integrable functions on $Q^\infty$.  The presentation of $\ch$ in this
manner is reminiscent of the presentation of a manifold in a preferred
manner, as for instance using a particular coordinate chart.

Can we give a reconstruction of $Q^\infty$ in an intrinsic way?  What new
structures are needed for  this purpose?

In the scheme we develop as a response to these questions, $Q^{\infty}$
emerges with its $C^{\infty}-$ structure only from certain states,
 {\em topology and differential features  being attributes of
particular classes of states and not universal properties of all states.}
Thus $Q^\infty$ emerges as a manifold only if the high energy components
in the observed states are strongly suppressed. When higher and higher
energies are excited, it gets more rough and eventually altogether ceases
to exist as a topological space modelled on a manifold.  Here by becoming
more rough we mean that $C^\infty$ becomes $C^K$
and correspondingly the $C^\infty-$ manifold $Q^\infty$ becomes a
$C^K-$ space $Q^K$.

\vglue 0.6cm
\sxn{ {\bf  What is Our Quantum System?}}
\vglue 0.5cm

The system we consider is that of a single particle, not that of quantum
fields or quantized strings.  The configuration space of a quantum field
has a very complex topology and its relation to our three-dimensional
spatial world is also nontrivial \cite{bdgmss1}.  Quantum
strings too share features
of
quantum fields.  For these reasons, we will not examine these systems.
For simplicity, we will also assume that the particle has no ``internal''
attributes like spin or flavor.

\vglue 0.6cm
\sxn{{\bf  Time is Special}}
\vglue 0.5cm

We have to assume that time evolution is given as a  unitary
operator $U(t)$ which is continuous in $t$.  Our analysis needs this
input. Time therefore persists
as an {\em a priori} irreducible notion even in our quantum approach.  It
would
be very desirable to overcome this limitation. (See \cite{CoRo} in this 
connection.)

There is more to be said on time, its role in measurement theory and as
the mediation between quantum and classical physics.  The appendix has
brief remarks on these matters.

It is true that in so far as our main text is concerned, $U(t)$ or the
Hamiltonian can be substituted by spatial translations, momenta or other
favorite observables.  But we think that time evolution is something
special, being of universal and  central interest to science.  It is for
this reason that we have singled out $U(t)$.

\vglue 0.6cm
\sxn{ {\bf  The Gel'fand-Naimark Theory}}
\vglue 0.5cm

The principal mathematical tool of our analysis involves this remarkable
theory \cite{fd} and, to some extent its developments in Noncommutative
Geometry \cite{co,col1,con,Landi,ma}.  We shall now give a crude and short
sketch of this
theory.

A $C^*$-algebra $\bar {\cc}$ with elements $c$ has the following
properties: a) It is an algebra over $\comm$. b) It is 
closed under an antinvolution $*$:

\be
*:c_j \in \bar {\cc} \Rightarrow c^{*}_{j} \in \bar \cc, ~c^{**}_{j} =
c_j,~
(c_1c_2)^*=c^{*}_{2}c^*_{1},~(\xi c_j)^*=\xi^*c^{*}_{j},
\ee

\noindent where $\xi$ is a complex number and $\xi^*$ is its complex
conjugate.
c) It has a norm $|| . || $ with the properties 
 $|| c^*|| = || c ||,~ || c^* c|| = || c ||^2$
for $ c \in \bar {\cc}$.
d) It is complete under this norm. (Hence the bar over $\cc$).  

A $*-$ representation $\rho$ of $\bar{\cc}$ on a Hilbert space $\ch$ is
the
representation of $\bar{ \cc}$ by a $C^*$-algebra of bounded operators on
 a  Hilbert space \cite{rs2} with the following features:
i)  The $*$ and norm for $\rho (\bar {\cc})$ are the operator adjoint
$\dagger$
and
operator norm (also denoted by $||\cdot ||$). ii)
$\rho(c^*)=\rho(c)^\dagger$.

$\rho$ is said to be a $*$-homomorphism because of ii).  We can also in a
similar manner speak of $*$-isomorphisms.  

We will generally encounter
$\bar{\cc}$ concretely as  an algebra of operators.  In any case, we will
usually omit the symbol $\rho$.

Note that a $*$-algebra (even if it is not $C^*$) is by definition closed
under an antinvolution $*$.

Let $\bar {\ca} $ denote a commutative  $C^*$-algebra. [There is no need
for now
to think of it as the closure of some $\ca$.]
Let $\{ x \}$ denote its space of inequivalent irreducible
$*$-representations
(IRR's) or its spectrum. [So $a\in \bar {\ca} \Rightarrow x(a) \in
\comm $.]
The Gel'fand-Naimark
theory then makes the following striking assertions: $\a)$ There is
a natural topology on $\{x\}$ making it into a Hausdorff topological
space
\cite{ency} $Q^0$. [We will denote the IRR's prior to introducing topology
by
$\{x\}$ and after doing so by $Q^\cdot$ with suitable superscripts.] 
$\b$) Let $\bar {\ca}_c$ be the $C^*$-algebra of~ $\comm$-valued
continuous
functions on $Q^0$.  Its $*$ is complex conjugation and its norm
$||\cdot ||$ is the supremum norm, $|| \phi ||= \sup_{x\in Q^{0}} |\phi
(x)|.$
Then $\bar {\ca}_c$ is $*$-isomorphic to $\bar {\ca}$.

We can thus identify $\bar{\ca}_c$ with $\bar{\ca}$, as we will often do.

The above results can be understood as follows.  By ``duality'', the
collection of $x(a)$'s for all $x$ defines a function $a_c$ on $\{ x\}$
by $a_c(x):=x(a)$. $ a_c$ is said to be the Gel'fand transform of $a$.

$\{x\}$ is
as
yet just a collection of points with no
topology.  How can we give it a natural topology? We want $a_c$ to be
$C^0$ in this topology.  Now the set of zeros of a continuous function is
closed.  So let us identify the set of zeros $C_a$ of each $a_c$ with a
closed set:

\be
C_a = \{x:x(a)\equiv a_c(x)=0\}.
\ee

\noindent The topology we seek is given by these closed sets.  The
Gel'fand-Naimark
theorem then asserts $\a$) and $\b$) for this topology, the isomorphism 
$\bar{\ca} \rightarrow \bar {\ca}_c$ being $a \rightarrow a_c$.

A Hausdorff topological space can therefore be equally well described by
a commutative  $C^*$ -algebra~ $\bar {\ca}$, presented for example using
generators.  That would be an intrinsic coordinate-free description of
the space and an alternative to using coordinate charts.

A $C^K$ - structure can now be specified by identifying an appropriate
subalgebra $\ca^K$ of $\bar {\ca} \equiv \ca^0$ and declaring that the
$C^K$ - structure is the one for which $\ca^K$ consists of $K$-times
differentiable functions.  [$\ca^K$ is a $*-$, but not a $C^*-$,
algebra for $K>0$.] The corresponding $C^K$-space is $Q^K$.
For $K=\infty$, we get the manifold $Q^\infty$.  We have the inclusions

\be
\ca^\infty \subset ... \subset \ca^K ... \subset \ca^0 \equiv \bar{\ca}
\ee

\noindent where

\be
\bar {\ca}^{(\infty)} = \bar{\ca}^{(K)} = \bar {\ca},
\ee

\noindent the bar as usual denoting closure.In contrast, $Q^\infty$ and 
$Q^K$ are all the same as sets,
being
$\{x\}$.

A dense $*$-subalgebra of a $C^*$-algebra $\bar{\cc}$ will be denoted
by
$\cc$ or $\cc^{\cdot}$, the superscript highlighting some additional
property.  The algebras $\ca^K$ are  examples of such $\cc^K$ for
$\bar{\cc}=\bar{\ca}$.  We will also reserve the symbols $\ca$ and
$\ca^\cdot$ for
the algebras  supposedly or otherwise leading to the classical
configuration space, the latter if it is $C^K$ being $Q^K$.

{\bf Example 1}: Consider the algebra $\ca$ generated by the
identity, an
element $u$ and its inverse $u^{- 1}$. Its elements are
$a= \sum_{N \in {\ZZ}} { \alpha_{N}u^{N}}$
where $\a_N$'s are complex numbers vanishing rapidly in $N$ at $\infty$.
The $*$
is defined by $u^*=u^{-1}, a^*=\sum a^{*}_{N} u^{-N}$.  As $\ca$ has
identity $\I1$ , there is a natural way to define inverse $a^{-1}$ too :
$a^{-1}$ is that element of $\ca$ such that $a^{-1}a=aa^{-1} = {\bf 1}$.
There
is also a canonical norm $||.||$ compatible with properties c)
\cite{co,con}:
$||a||$ = Maximum of $|\lambda|$ such that $a^*a-|\lambda|^2$ has no
inverse.

The space $Q^\infty$ for this $\ca$ is just the circle $S^1$, $u_c$
being the function with value $e^{i\theta}$ at $e^{i\theta} \in S^1$.

If similarly we consider the algebra associated with $N$ commuting
unitary
elements, we get the $N$-torus $T^N$.  If  for $N=2$, the generating
unitary elements do not commute, but fulfill  $u_1u_2= \omega u_2 u_1,
\omega$ being any phase, we get the noncommutative torus
\cite{cor,co}. It is the ``rational'' or ``fuzzy'' torus if $\omega^K=1$
for
some $K \in
{\ZZ}$, otherwise it is ``irrational'' \cite{ma,bfss}.

\vglue 0.6cm
\sxn{{\bf  What Time Evolution Tells Us}}
\vglue 0.5cm

Let $\ch^0\equiv \bar {\ch}^0$ denote the Hilbert space of state vectors.
[The bar has been introduced as usual to emphasize norm-completeness.]
The
unitary operator $U(t)$ is defined on all of $\ch^0$ and is continuous
there.  Let us assume that $U(t)$ has a discrete spectrum with
orthonormal  eigenvectors $\phi_n$ spanning $\ch^0$:

\be
U(t) \phi_n=e^{-iE_{n}t}\phi_n,~(\phi_n,\phi_m)= \delta_{nm}, n\in {\IN
}.
\ee

The completeness of $\{\phi_n\}$ means that every $\psi^0 \in \ch^0$ has
the expansion

\be
\psi^0= \sum_{n \in {\IN}} a^{0}_{n}\phi_n~~{\rm with}~~\sum_{n\in {\IN}} 
|a^{0}_{n}|^2=(\psi^{0},\psi^{0})< \infty.
\ee

\noindent The time evolution of $\psi^0$ is

\be
U(t)\psi^0=\sum a^{0}_{n} e^{-iE_{n}t} \psi^{0}.
\ee

But $U(t)$ is not always differentiable on all of $\ch^0$. That requires
that
$a^{0}_{n}$ fulfill $\sum|a^{0}_{n}|^2E^{2}_{n}<\infty.$
[Otherwise
the norm of $\sum a^{0}_{n} \frac {d}{dt} (e^{-iE_{n}t} \psi^0)$
diverges.]
 Let $a^{1}_{n}$ denote these $a^{0}_{n}$'s, $\psi^1$'s the corresponding
vectors and $\ch^1$ the subspace of these vectors.  Then
\be
\ch^1 = \{\psi^1 \in \ch^0: \psi^1=\sum a^{1}_{n} \phi_n 
~{\rm with}~
\sum|a^{1}_{n}|^2|E^{2}_{n}|<\infty~\},
\ee

\be
~~\frac {dU(t)}{dt} \psi^1 = -i\sum a^{1}_{n} e^{-iE_{n}t} \phi_n. 
\ee

\noindent $\ch^1$ is a dense subspace of $\ch^0$:

\be
\ch^1 \subseteq \ch^0,~~\bar{\ch}^1 = \ch^0 \equiv \bar {\ch}^0.
\ee

\noindent It is the domain \cite{rs2} of the Hamiltonian $H$, the latter
being defined by $H\psi^1=-i\frac{dU(t)}{dt} \psi^{1}|_{t=0}$.

It could of course happen that $\ch^1=\ch^0$. 
That would be the case if $\ch^0$ is finite dimensional, or if $|E_n|$
does not diverge as  $n\rightarrow \infty$ (that is if $H$ is bounded
\cite{rs2}). In either case, there is no classical underlying manifold in
our
approach, {\em unbounded operators being of essential significance for
the recovery
of classical attributes}.  So we consider only systems not covered by the
above possibilities.  Then we have the strict inclusion $\ch^1 \subset
\ch^0$.

The subspace $\ch^K$ where $U(t)$ is $K$-times differentiable is found in
a similar way,

\be
\ch^K = <\psi^K \in \ch^0:\psi^K=\sum a^{K}_{n} \phi_n, \sum
|a_{n}^{K}|^2|E_n|^M<\infty~~{\rm for~ all}~~ M\in \{0,1,\ldots,K\}>,
\ee

\noindent while

\be
\ch^\infty = <\psi^\infty \in \ch^0: \psi^\infty=\sum a^{\infty}_{n}
\phi_n, \sum |a^{\infty}_{n}|^2 |E_n|^M<\infty ~~{\rm for~ all} ~~M \in
\IN>.
\ee

\noindent We have

\be
\ch^\infty \subset \ldots \subset \ch^K \subset \ldots \subset
\ch^0\equiv \bar {\ch}^0. 
\ee

{\em Note how higher and higher energies are suppressed in $\ch^K$ as we
go up
in $K$}.

{\bf Example 2:} For a nonrelativistic particle on ${\bf R}^3$  with
Hamiltonian $H=-\frac{1}{2m} \vec\bigtriangledown ^2 +a$ a smooth bounded
potential,  $\ch^\infty = C^\infty ({\bf R}^3) \cap L^2 ({\bf
R}^3)$.

\vglue 0.6cm
\sxn{ {\bf  Axioms on Probability Densities} }
\vglue 0.5cm

Let us first note a few heuristic results.

Let $W$ be a commutative $*$-algebra of bounded operators on $\ch^0$. 
Let $\{x\}$  be its spectrum, $|x>$ the
corresponding states (assumed nondegenerate) in $\ch^0$ and $I=\int
\omega|x><x|$ the
resolution of identity. 

For $\psi^0,~\chi^0 \in \ch^0$, we can now define the analogue of a  
probability
density, a hermitean form $\psi^{0 ~\dagger}\chi^0 $ which is a function on
$\{x\}$, by

\be
\psi^{0 ~\dagger}\chi^0 (x) = <\psi^0|x><x|\chi^0>.
\ee

It is $L^1$ since

\be
\int \omega\psi^{0 ~\dagger}\chi^0 (x) = (\psi^0, \chi^0).
\ee

We now come to our central axioms for the recovery of classical topology
from quantum physics.  {\em They are}, in so far as we can tell, {\em
new}, {\em
to be
added on to existing quantum principles whenever we desire an emergent
classical topology}.

Let $\cb$ be a *-algebra with properties a) to c) of Section 6, spectrum $Q^\cb$ 
with points
$x$, corresponding state vectors $|x>$ (assumed non-degenerate) and
resolution of unity $I=\int \omega |x><x|$.

We concentrate now on $\ch^\infty$.  For $\psi^\infty, ~\chi^\infty$ in
$\ch^\infty$, we have then an integrable function
$\psi^{\infty\dagger}\chi^\infty$ on $\{x\}$ as in the above construction.
It will not be in the
Gel'fand
transform $\cb_c$ of $\cb$ for an arbitrary choice of $\cb$.

\vglue 0.6cm
\noindent {\em {\bf The Axioms}}
\vglue 0.5cm
\begin{itemize}
\item A1.  There exists at least one choice $\ca^\infty$ for $\cb$  such that
$\psi^{\infty \dagger}\chi^{\infty}$ as constructed above is in $\ca^\infty$ 
and generates
it. [ We are here identifying  $\ca^\infty$ with $\ca^{\infty}_c$.]

\item A2.  (Locality): $\ca^\infty$ is {\em local}, that is, there exists an $M
\in \IN^+$ such that

\be
[a_K, [a_{K-1}, \ldots [a_0, H] \ldots ] = 0 ~~{\rm for}~~\forall a_i \in
\ca^{(\infty)} ~~{\rm and}~~ K \geq M.\label{8.1}
\ee

\item  A3. If the choice of $\ca^\infty$ with properties A1,2 is
not unique for
a particular system, it does become unique when a sufficiently large
number of systems are considered, $\ca^\infty$ being the common algebra
to be found among all of them.

\item A4.  The classical configuration space as a topological space is the
spectrum $Q^0$ of $\bar {\ca}^\infty=\bar{\ca}$ whereas it is given as a
manifold $Q^\infty$ by treating $\ca^\infty$ as $C^\infty$-functions.
\end{itemize}
\vglue 0.6cm
\noindent {\em {\bf Explanations}}
\vglue 0.5cm
\begin{itemize}
\item  A1: For example 1, $\ca^\infty$ is just $C^\infty({\bf
R}^3) \cap L^1 ({\bf R}^3)$.

\item   A2: Equation  (\ref{8.1}) with $K=M$ of course implies its validity for all $K>M$.

For a particle on $S^1$ say, with A1 alone,
$\ca^\infty$ can be smooth functions on {\em either} $S^1$ {\em or}
$\ZZ$
(the
quantum momentum space). A2 attempts to sort out this sort of ambiguity
since
interactions generally are local only on configuration space.  When the
latter is a standard manifold, the meaning of A2 is that $H$ must be a
differential operator of finite order.  The {\em order} of $H$ then is
just the {\em least} value of $K$.  We shall adopt this always as our
definition of order of $H$:  {\em It is the least value of $K$ for which
{\rm (\ref{8.1})} is true}.

\item A3:  The algebra $\ca^\infty$ may not be fixed even with A2.   For a free
particle on $S^1$, we still have the  possibility of considering
functions on $S^1$ or ${\ZZ}$ as $\ca^\infty$.  The intention of A3 is
to
resolve this ambiguity by considering several interactions, the
assumption being that they will always be local on configuration space,
but not so on other spaces.

The Hamiltonian $[{\underline{p}}^2+m^2]^{\frac {1}{2}}$ for a
relativistic
particle of mass $m$ [$\vec{p}$ = momentum] does not fit in our
scheme, being nonlocal on configuration space.  It is also generally
rejected  in the presence of interactions for this nonlocality, so perhaps
we
need not become anxious thinking of this operator.

In any case, a better principle than our locality to fix $\ca^\infty$
uniquely would be desirable.

\item A4: $I$ fixes only $\omega |x><x|$, and not $\omega$ and $|x><x|$
separately. This fact creates uncertainties in the definition of
$\psi^{\infty ~\dagger} \chi^\infty$.  One way to resolve this uncertainty 
adequately is
to choose  $\omega$ as some particular  smooth volume form  $\omega_0$
on $Q^\infty$, thereby defining $\qquad|x><x|$.
Equivalently,  we can fix a smooth volume form $\omega_0$ on $Q^\infty$
and
define $\psi^{\infty ~\dagger} \chi^\infty$ by the equality 
$\omega <\psi^\infty|x><x|\chi^\infty> =
\omega_0\psi^{\infty ~\dagger}\chi^\infty(x)$.  Alternatively, we can modify 
A1 to the
requirement that there exists a choice $\ca^\infty$ for $\cal {B}$ such
that the set of $\omega \psi^{\infty ~\dagger} \chi^\infty$ forms an $\cal
{A}^\infty$-module. 
\end{itemize}
\vglue 0.6cm
\noindent {\bf Further Remarks}
\vglue 0.5cm

Once $\ca^\infty$ has been fixed, we can construct the  functions
$\psi^{K^\dagger} \chi^K$ from $\psi^K,~\chi^K \in \ch^K$.  With
decreasing $K$,
they
would give vectors which are fewer and fewer times differentiable.
Eventually for
$K=0$, they would all be defined only in the $L^1$-sense.  Thus when
higher and higher energies are increasingly excited, things fall apart and
the  topology of
configuration space increasingly gets rougher. It  disappears
altogether as
a topological space modeled on a manifold when $K=0$ is reached.  This
interesting point was
first discussed in ref. 2.

This comment can be rephrased in terms of modules of forms as in the
comments above on A4.

The tie-up between $\ch^\infty$ and $\ca^\infty$ in our approach is not
prompted by random fancy.  It is this connection that links time
evolution to classical spatial topology.

Incidentally it can checked in standard examples, as for instance the
commutative tori of example 1, that $\ch^\infty$ are
$\ca^\infty$ - modules.

\vglue 0.6cm
\sxn{ {\bf  Dimension and Metric}}
\vglue 0.5cm

The topological space $Q^\cdot$ constructed with the level of generality
maintained hitherto need not even resemble a manifold for our limited
axioms. Further
conditions like those discussed by Connes \cite{con} are necessary to
avoid what  
a
classical physicist may consider as pathologies.

Suppose that such conditions are also met and that $Q^\infty$ is a
manifold.  We can then find its dimension in the usual way.

There is also a novel manner to find its dimension $d$ from the spectrum
$\{{\lambda}_n\}$
of $H$: If $H$ is of order $N$, $|{\lambda}_n|$ 
grows like $n^{N/d}$ as $n \rightarrow \infty$ \cite{co,col1,con,fg}.

We can find a metric as well for  $Q^\infty$ \cite{co,con,fg}:  It is
specified by
the
distance 

\be
d(x,y)=\{\sup_{a}|a_c(x)-a_c(y)|:\frac {1}{N!} ||\underbrace {[a,[a,\ldots
[a,H]
\ldots ]}_{N ~a's}||\leq 1\}.
\ee

\noindent This remarkable formula gives the usual metric for the Dirac
operator
$[N=1]$ \cite{co,col1} and  the Laplacian $[N=2]$ \cite{fg}.

\vglue 0.6cm
\sxn{{\bf What is Quantum Topology?}}
\vglue 0.5cm

A question of the following sort often suggests itself when encountering
discussions of topology in quantum gravity:  If $Q$ is a topological
space, possibly with additional differential and geometric structures
[``classical'' data], what is meant by {\em quantizing} $Q$?

It is perhaps best understood as: {\em finding an algebra of operators on
a Hilbert space from which $Q$ and its attributes can be reconstructed}
[much as in the Gel'fand-Naimark theorem].

\vglue 0.6cm
\sxn{{\bf Topology Change}}
\vglue 0.5cm

We now use the preceding ideas to discuss topology change, following
ref. 2. [See ref. \cite{ma2} for related work.]

There are indications from theoretical considerations that spatial
topology in quantum gravity cannot be a time-invariant attribute, and
that its transmutations must be permitted in any eventual theory.

The best evidence for the necessity of topology change comes from the
examination of the spin-statistics connection for the so-called geons
\cite{fs,sor,bmss}.  Geons are solitonic excitations caused by twists in
spatial
topology.  In the absence of topology change, a geon can neither
annihilate nor be pair produced with a partner geon, so that no geon has
an associated antigeon.

Now spin-statistics theorems generally emerge in theories admitting
creation-annihilation processes \cite{bdgmss1,sor,ds}.  It can therefore be
expected to
fail for geons in gravity theories with no topology change.  Calculations
on geon quantization in fact confirm this expectation \cite{sor,abbjrs}.

The absence of a universal spin-statistics connection in these gravity
theories is much like its absence for a conventional nonrelativistic
quantum particle which too cannot be pair produced or annihilated.   Such
a particle can obey any sort of statistics including parastatistics
regardless of its intrinsic spin.  But the standard spin-statistics
connection can be enforced in nonrelativistic dynamics also by
introducing
suitable creation-annihilation processes \cite{see}.

There is now a general opinion that the spin-statistics theorem should
extend to gravity as well.  Just as this theorem emerges from even
nonrelativistic physics once it admits pair production and annihilation
\cite{bdgmss1}, quantum gravity too can be expected to become compatible
with this
theorem after it allows suitable topology change \cite{ds}.  In this
manner,
 the desire for the
usual spin-statistics connection leads us to look for a quantum gravity
with transmuting topology.

Canonical quantum gravity in its elementary form is predicated on the
hypothesis that spacetime topology is of the form $\Sigma \times {\bf
{R}}$
(with
${\bf {R}}$ accounting for time) and has an eternal spatial topology.
This
fact has led to numerous
suggestions that conventional canonical gravity is inadequate if not
wrong, and must be circumvented by radical revisions of spacetime
concepts \cite{blms}, or by improved approaches based either on functional
integrals and cobordism \cite{ds} or on alternative quantization methods.

Ideas on topology change were first articulated in quantum gravity, and
more specifically in attempts at semiclassical quantization of classical
gravity.  Also it is an attribute intimately linked to gravity in the
physicist's mind.  These connections and the apparently revolutionary
nature of topology change as an idea have led to extravagant speculations  
about twinkling topology in quantum gravity and their impact
on fundamental concepts in physics.

Here we show that models of quantum particles exist which admit topology
change or contain states with no well-defined classical topology.  {\em
This is so even  though gravity does not have a central role in our ideas
and is significant only  to  the extent that metric is important for a
matter Hamiltonian.}
These models use only known physical principles and have no
revolutionary content, and at least suggest that topology change in
quantum gravity too may be achieved with a modest physical input and no
drastic alteration of basic laws.

We consider particle dynamics as usual.
 The configuration space of a particle being ordinary space, we are thus
imagining a physicist probing spatial topology using a particle.

Let us consider a particle with no internal degrees of freedom living on
the union $Q^{\prime}$ of two intervals which are numbered as 1 and 2:

\be
Q^{\prime}=[0,2\pi] \bigcup [0,2\pi] \equiv Q^{\prime}_1 \bigcup
Q^{\prime}_2~.\label{11.1}
\ee

\noindent It is convenient to write its wave function $\psi$ as $(\psi_1,
\psi_2)$,
where each $\psi_i$ is a function on $[0,2\pi]$ and $\psi^*_i \psi_i$ is
the probability density on $Q^{\prime}_i$. The scalar product between
$\psi$ and another wave function $\chi=(\chi_1,\chi_2)$ is

\be
(\psi,\chi)=\int_0^{2 \pi} dx \sum_i (\psi^*_i\chi_i)(x)~.
\label{11.2}
\ee

It is interesting that we can also think of this particle as moving on
$[0,2\pi]$ and having an internal degree of freedom associated with the
index $i$.

After a convenient choice of units, we define the Hamiltonian formally by

\be
(H \psi)_i(x)=-\frac{d^2 \psi_i}{dx^2}(x)~ \label{11.3}
\ee

\noindent [where $\psi_i$ is assumed to be suitably differentiable in the
interval
$[0,2\pi]$]. This definition is only formal as we must also specify its
domain
$\ch^1$
\cite{rs2}. The latter involves the statement of the boundary conditions
(BC's) at $x=0$ and $x=2\pi$.

Arbitrary BC's are not suitable to specify a domain: A symmetric operator
${\cal O}$
with domain $D({\cal O})$ will not be self-adjoint unless the following
criterion is also fulfilled:

\be
{\cal B}_{\cal O}(\psi,\chi)\equiv (\psi,{\cal O} \chi)-(
{\cal O}^{\dagger}\psi,\chi)=0~~{\it for~all}~\chi \in D({\cal O})
 \Leftrightarrow
\psi \in D({\cal {O}})~.
\label{11.4}
\ee

For the differential operator $H$, the form ${\cal B}_{H}( \cdot, \cdot)$
is
given by

\be
{\cal B}_H(\psi,\chi)=
\sum_{i=1}^{2} \left [ -\psi^{*}_{i}(x) 
\frac{d \chi_{i}(x)}{dx}+
\frac{d\psi^{*}_{i}(x)}{dx}\chi_{i}(x)\right ]^{2 \pi}_{0}~.\label{11.5}
\ee

\noindent It is not difficult to show that there is a $U(4)$ worth of
$D(H)\equiv\ch^1$ here
compatible with (\ref{11.4}).

We would like to restrict this enormous choice for $D(H)$, our intention
not being to study all possible domains for $D(H)$. So let us 
restrict ourselves to the domains

\be
\ch^{(1)}_{u}=\{\psi \in C^2(Q^{\prime}):\psi_i(2 \pi) =
u_{ij}\psi_j(0),~~
\frac{d\psi_{i}}{dx}(2\pi)=u_{ij}
\frac{d\psi_{j}}{dx}(0),~u \in U(2)\}~. \label{11.6}
\ee

\noindent  These domains have the
virtue
of being compatible with the definition of momentum in the sense discussed
in ref. 2.

There are two choices of $u$ which are of particular interest:

\be
a)~~~~~~~~~~
 u_{a}=\left( \begin{array}{cc}
0 & e^{i\theta_{12}} \\
e^{i\theta_{21}} & 0
\end{array}\right) , \label{11.7}
\ee

\be
b)~~~~~~~~~
 u_{b}=\left( \begin{array}{cc}
e^{i\theta_{11}} & 0 \\
0 & e^{i\theta_{22}}
\end{array} \right) .\label{11.8}
\ee

In case $a$, the density functions $\psi^*_i \chi_i$ fulfill
\be
(\psi^*_1 \chi_1)(2\pi)=(\psi^*_2 \chi_2)(0)~,
\ee

\be
(\psi^*_2 \chi_2)(2 \pi)=(\psi^*_1 \chi_1)(0)~.\label{11.9}
\ee

\noindent Figure 1 displays (\ref{11.9}), these densities being the same at the
points
connected by broken lines.

  \begin{figure}[hbtp]
  \epsfysize=4cm
  \epsfxsize=6cm
 \centerline{\epsfbox{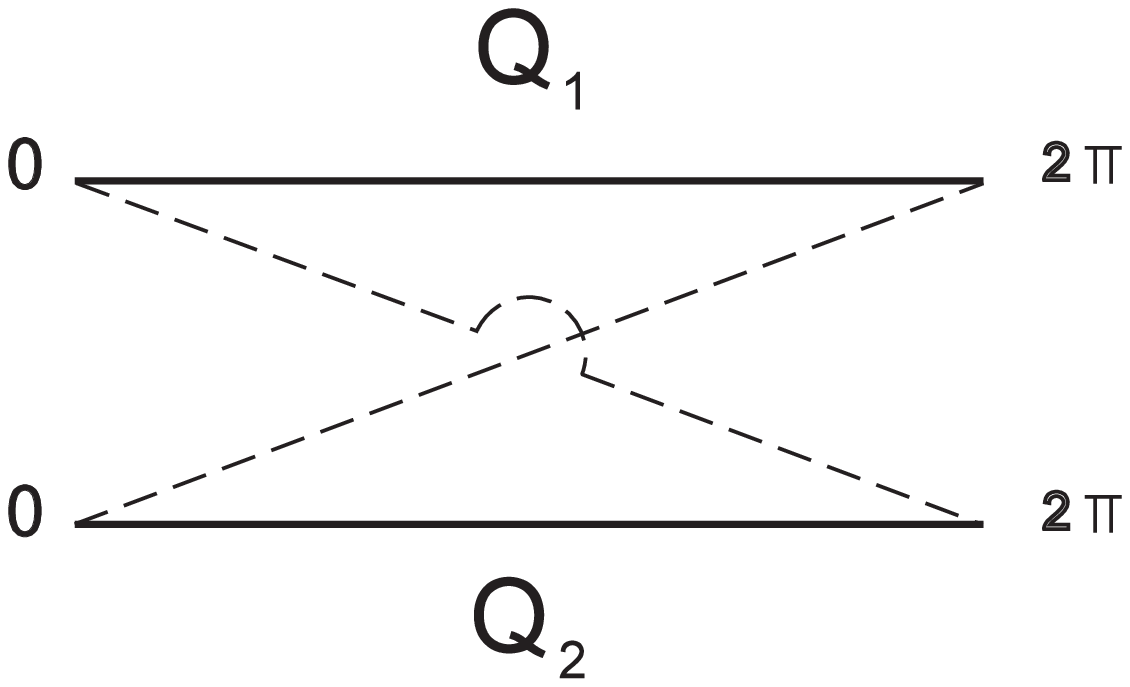}}
 \caption{  In case $a$, the density
functions are the same at the points joined by broken lines in this Figure.}
\end{figure}

In case $b$, they fulfill, instead,
\be
(\psi^*_1 \chi_1)(2\pi)=(\psi^*_1 \chi_1)(0)~,
\ee

\be
(\psi^*_2 \chi_2)(2 \pi)=(\psi^*_2 \chi_2)(0)~\label{11.10}
\ee

\noindent which fact is shown in a similar way in Figure 2.

\begin{figure}[hbtp]
  \epsfysize=4cm
  \epsfxsize=6cm
 \centerline{\epsfbox{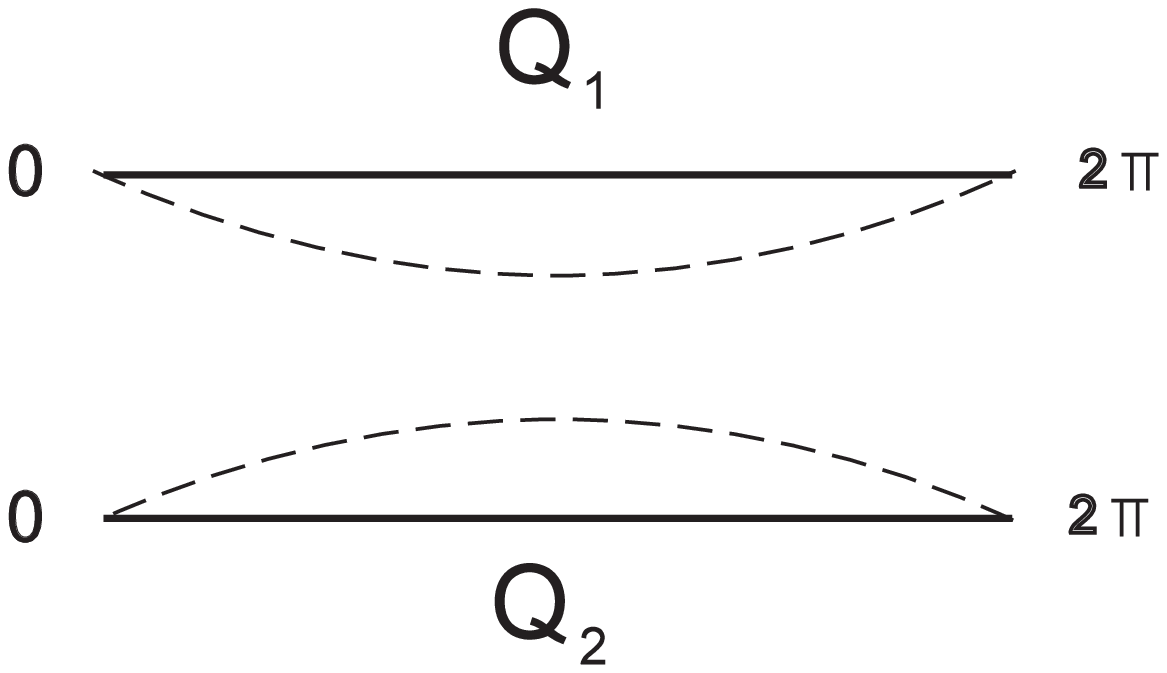}}
 \caption{ In case $b$, the density
functions are the same at the points joined by broken lines in this Figure.}
\end{figure}

Continuity properties of $\psi^*_i \chi_i$ imply that we can identify the
points joined by dots to get the classical configuration space $Q$. It is
\underline{not} $Q^{\prime}$, but rather a circle $S^1$ in case $a$ and
the union $S^1 \bigcup S^1$ of two circles in case $b$.

The requirement $H^M \ch^\infty \subset \ch^\infty$ for all 
$M\in {\IN}$ means just that
arbitrary derivatives of $\psi^*_i \chi_i$ are continuous at the points
joined by broken lines, that is on $S^1$ and $S^1 \bigcup S^1$ for the two
cases.
We can prove this easily using (\ref{11.6}).
In this way we also recover $S^1$ and $S^1\bigcup S^1$ as manifolds.

When $u$ has neither of the values (\ref{11.7})  and (\ref{11.8}), then $Q$
becomes the union of two intervals. The latter
happens for example for

\be
u=\frac{1}{\sqrt 2}\left( \begin{array}{cc}
1 & 1 \\
-1 & 1
\end{array}\right)~~. \label{11.11}
\ee

\noindent In all such cases, $Q$ can be regarded as a manifold with
boundaries as shown
by the argument above.

Summarizing, we see that the character of the underlying classical
manifold depends on the domain of the quantum Hamiltonian and
can change when $u$ is changed.

It is possible to reduce the $u$ in the BC to $\bf 1$ by introducing a
connection. Thus since $U(2)$ is connected, we can find a $V(x)\in U(2)$
such that

\be
V(0)={\bf 1},~~~~V(2 \pi)=u^{-1}.\label{11.12}
\ee

\noindent Using this $V$, we can unitarily transform $H$ to the new
Hamiltonian
\be
H^{\prime}=VHV^{-1},
\ee
\be
(H^{\prime}\psi)_i(x)=-\left[\frac{d}{dx}+A(x)\right]_{ij}^{2}\psi_j(x)~,
\ee

\be
A(x)=V(x)\frac{d}{dx}V^{-1}(x)~.  \label{11.13}
\ee

\noindent With suitable physical interpretation, the system defined by
$H^{\prime}$
and the domain

\be
D_{\bf 1}(H^{\prime})=V D_u(H)\equiv \{ \phi:~\phi=V\psi,~~\psi\in
\ch^{1}_{u}\}
\label{11.14}
\ee

\noindent is evidently equivalent to the system with Hamiltonian $H$ and
domain
$\ch^{1}_{u}$. Note in this connection that density functions on
$Q_i^{\prime}$
are $\psi_i^*\chi_i$ and not $(V\psi)_i^*(V\chi)_i$.

\vglue 0.6cm
\noindent {\em \bf  Dynamics for Boundary Conditions}
\vglue 0.5cm

We saw in the previous section that topology change can be achieved in
quantum physics by treating the parameters in the BC's as suitable
external parameters which can be varied. 

However it is not quite satisfactory to have to regard $u$ as an external
parameter and not subject it to quantum rules.  We now therefore promote
it to an operator, introduce its conjugate variables and modify the
Hamiltonian as well to account for its dynamics.  The result is a closed
quantum system.  It has no state with a sharply defined $u$.  We cannot
therefore associate one or two circles with the quantum particle and
quantum spatial topology has to be regarded as a superposition of
classical spatial topologies.  Depending on our choice of the
Hamiltonian, it is possible to prepare states where topology is peaked at
one or two $S{^1}$'s for a long time, or arrange matters so that there is
transmutation from one of these states to another.

Dynamics for $u$ which determines BC's is best introduced in the
connection picture where the domain of $H^{\prime}$ is associated with
$u=1$. We assume this representation hereafter.

Quantization of $u$ is achieved as follows. Let $T(\alpha)$ be the
antihermitean generators of the Lie algebra of $U(2)$ [the latter being
regarded as the group of
$2
\times 2$ unitary matrices] and normalized according to $Tr~ T(\alpha)
T(\beta)=-N \delta_{\alpha \beta}$, $N$ being a constant. Let $\hat u$ be the
matrix of quantum operators representing the classical $u$. It fulfills

\be
\hat u_{ij} \hat u^{\dagger}_{ik} = {\bf 1} \delta_{jk},~~[\hat u_{ij}, \hat
u_{kh}]=0~, \label{11.15}
\ee
$\hat u_{ik}^{\dagger}$ being the adjoint of $\hat u_{ik}$.
The  operators conjugate to $\hat u$ will be denoted by $L_{\alpha}$. If
\be
[T_{\alpha},T_{\beta}]=c_{\alpha \beta}^{\gamma}T_{\gamma},
\ee

\be
c_{\alpha \beta}^{\gamma}={\rm structure~constants~of}~U(2),
\label{11.16}
\ee
$L_{\alpha}$ has the commutators
\be
[L_{\alpha},\hat u]=-T(\alpha)\hat u~, 
\label{11.16.5}
\ee

\be
[L_{\alpha},L_{\beta}]=c_{\alpha\beta}^{\gamma}L_{\gamma}, \label{11.17}
\ee

\be
[T(\alpha)\hat u]_{ij}\equiv T(\alpha)_{ik}\hat u_{kj}.
\ee

If $\hat V$ is the quantum operator for a function $V$ of $u$,
$[L_{\alpha},\hat V]$ is
determined by (\ref{11.13}) and (\ref{11.17}). 

The Hamiltonian for the combined particle-$u$ system can be taken to be,
for example,
\be
\hat H = \hat H^{\prime} + \frac{1}{2I}\sum_{\alpha} L^2_{\alpha},
\ee

\be
\hat H ^{\prime}=-\left( \frac{d}{dx}+ \hat A (x)\right)^2,~~\hat
A(x)\equiv \hat V(x) \frac{d}{dx} \hat V^{-1}(x), \label{11.18}
\ee

\noindent $I$ being the moment of inertia.

Quantized BC's with a particular dynamics are described by (\ref{11.15}),
(\ref{11.16.5}),(\ref{11.17}) and (\ref{11.18}).

The general state vector in the domain of $\hat H$ is a superposition of
state vectors $\phi \otimes_{\bf C}|u\rangle $ where $\phi \in D_{\bf
1}(H^{\prime})$ and $|u\rangle$ is a generalized eigenstate of $\hat u$:

\be
\hat u_{ij}|u\rangle=u_{ij}|u\rangle,~~\langle u^{\prime}|u
\rangle=\delta(u^{\prime -1}u)~.\label{11.20}
\ee
The $\delta$-function here is defined by

\be
\int du f(u)\delta(u^{\prime -1}u)=f(u^{\prime}),\label{11.21}
\ee

\noindent $du$ being the (conveniently normalized) Haar measure on $U(2)$.
Also

\be
\hat A(x) |u\rangle = A(x) |u \rangle~.\label{11.22}
\ee

It follows that the classical topology of one and two circles is
recovered on the states $\sum_\lambda  C_{\lambda}
\phi^{(\lambda)}\otimes_{\bf C}|u_{a}\rangle $ and
$\sum_\lambda
D_{\lambda}\phi^{(\lambda)}\otimes_{\bf
C}|u_b\rangle,~[C_{\lambda},~D_{\lambda}\in
{\bf C}, ~\phi^{(\lambda)}\in D_{\bf 1}(H^{\prime})$] with the two fixed
values $u_a$,  and $u_b$ of (\ref{11.7}) and (\ref{11.8}) for $u$.

But these are clearly idealized unphysical vectors with infinite norm.
The best we can do with normalizable vectors to localize topology around
one or two circles is to work with the wave packets

\be
\int du f(u) \phi \otimes_{\bf C} |u\rangle~,
\ee

\be
\int du |f(u)|^2 < \infty \label{11.23}
\ee

\noindent where $f$ is sharply peaked at the $u$ for the desired topology.
The
classical topology recovered from these states will only approximately be
one or two circles, the quantum topology also containing admixtures from
neighboring topologies of  two intervals.

A localized state vector of the form (\ref{11.23}) is not as a rule an
eigenstate of a Hamiltonian like $\hat H$. Rather it will spread in
course of time so that classical topology is likely to
 disintegrate
mostly into that of two intervals.  We can of
course localize it around one or two $S^1$'s for a very long time by
choosing $I$ to be large, the classical limit for topology being achieved
by letting $I \rightarrow \infty$. By adding suitable potential terms, we
can also no doubt arrange matters so that a wave packet concentrated
around $u=u_a$ moves in time to one concentrated around  $u=u_b$. This
process would be thought of as topology change by a classical physicist.

The preceding considerations on topology change admit generalizations to
higher dimensions as explained in ref.2.

\vglue 0.6cm
\sxn{ {\bf  Final Remarks}}
\vglue 0.5cm

In this essay, we have touched upon several issues concerning quantum
topology and showed their utility for research of current interest such
as topology change and fuzzy topology.  Our significant contribution, if
any, here has been in formulating new fundamental problems with
reasonable clarity.  We have also sketched a few answers, but they are
tentative and incomplete.

\vglue 0.6cm
\sxn{{\bf Acknowledgments}}
\vglue 0.5cm

The work reported in this article is part of an ongoing program with
several colleagues.  I have especially benefited from discussions with
Jan Ambjorn,Peppe Bimonte, T.R. Govindarajan, Gianni Landi, Fedele Lizzi,
Beppe Marmo,
Shasanka Mohan Roy, Alberto Simoni and Paulo Teotonio-Sobrinho in its
preparation.I am also deeply grateful to Arshad Momen for his 
extensive and generous help in the preparation of this paper.  This work 
was supported by the U.S. Department of Energy
under Contract Number DE-FG-02-85ER40231.

\vglue 0.6cm
\appendix
\sxn{Appendix}
\vglue 0.5cm

Standard quantum measurement theory deals with instantaneous measurements
which are then shown to be capable of observing only commuting sets of
operators.  We assume these observables  to be bounded, closed under
$\dagger$ and
complete
also.  They generate a commutative  normed $*$-algebra $\cb$, the norm
$||\cdot
||$ being the operator norm and $*$ being $\dagger$.  We can recover a
Hausdorff
topological space from $\cb$ (or from its $C^*$-completion $\bar{\cb}$).
In this manner, we can recover the classical configuration space from the
instantaneous measurement of a correct commuting set.

In the text, we were not overtly concerned with instantaneous or any other
sort of measurements.  Rather we were concerned with the construction of a
correct algebra for the recovery of the classical configuration space.
The reason why this algebra should be adapted to $\ch^\infty$ was
explained there.  In that discussion, we required that the hermitean form
$\psi^{\infty~ \dagger}\chi^\infty (\psi^\infty, \chi^\infty \in \ch^\infty)$ should generate a {\em
commutative}}
algebra and then tried to devise rules to pick out the right form and its
algebra $\ca^\infty$.  Our insistence on this commutative nature came partly
from the desire to reconstruct a Hausdorff space, as is appropriate for
classical physics.  Once we have this $\ca^\infty$, it can serve as our
choice of $\cb$.

We thus see that there is a link between quantum instantaneous
measurements and classical topology and that this link is mediated by
$\ch^{\infty}$ and $\ca^\infty$.

There is a simple argument based on temporal continuity of experimental
outputs which confirms the standard analysis that instantaneous
measurements can be made only on {\em commuting} sets.

Suppose that we first measure an observable $a (= a^\dagger)$ and
then
an observable $b(=b^\dagger)$ an infinitesimal time $\delta$ later.  Let us for
simplicity assume that $a$ and $b$ have discrete spectra.  We consider two
cases in turn.

\vglue 0.6cm
\noindent {\bf Case 1.} ~$ab \neq ba$.
\vglue 0.5cm

In this case, if $|\alpha>, |\beta>$ are eigenvectors of norm 1 for
eigenvalues $\alpha,\beta$ of $a$ and $b$ (assumed nondegenerate for
simplicity), the probability of finding $\alpha$ for $a$ in the state
vector  
$|\psi>(<\psi|\psi>=1)$ is $|<\alpha|\psi>|^2$.  The vector after finding
$\alpha$ is $|\alpha>$. The probability of finding
$\beta$ for $b$ an instant later is thus $|<\beta|\alpha>|^2$.  The
probability $P(\beta,\alpha)$ of finding $\alpha$ for $a$ and then $\beta$
for $b$ in the limit $\delta \downarrow 0$ is the product of these
probabilities:

\be
P(\beta,\alpha)=|<\beta|\alpha>|^2~|<\alpha|\psi>|^2.
\ee

\noindent As this is not symmetric in $a$ and $b$, the order of the
measurements of
$a$ and $b$ will give different answers, even if their time separation is
negligible. So insistence on time-continuity excludes the possibility of 
their simultaneous
measurement.

\vglue 0.6cm
\noindent {\bf Case 2.~~ $ab=ba$}
\vglue 0.5cm

Let us suppose that $a$ and $b$ form a complete commuting set and let
$|\alpha^\prime,\beta^\prime>$ denote the simultaneous eigenvectors of
$a$
and $b$
for eigenvalues $\alpha^\prime$ and $\beta^\prime$,
 with
$<\alpha^{ \prime \prime},\beta^{ \prime \prime}| \alpha^\prime,
\beta^\prime > =
\delta_{\alpha^{ \prime \prime} \alpha^{\prime}}
\delta_{\beta^{ \prime \prime} \beta^{\prime} }$.  Then 
the probability of finding $\alpha$ for $a$ in $|\psi>$ is 
$P(\alpha) = 
\sum_{\beta^{\prime}} |<\alpha, \beta^\prime | \psi>|^2$.  The state
vector after
finding $\alpha$ is $\sum_{\beta^\prime} 
\frac
{1}{\sqrt{P(\alpha)}}|\alpha,
\beta^\prime > <\alpha, \beta^\prime |\psi>$.  The probability of finding
$\beta$ in
this state as $\delta \downarrow 0 $ is $\frac
{|<\alpha,\beta|\psi>|^2}{P(\alpha)}$. Hence now 

\be
P(\beta,\alpha) = |<\alpha, \beta| \psi >|^2.
\ee

\noindent This is symmetric in $a$ and $b$, so time-continuity does not
exclude
their measurements within a vanishingly small temporal separation.

Instantaneous measurements are linked not just to classical topology, they
are linked to classical physics in yet another way:  We have emphasised that
they
observe only
commutative  algebras $\cb$ [we continue to assume that $\cb$ is of the sort 
indicated
earlier],but the state $|\psi><\psi|$ restricted to $\cb$ is just a
classical probability distribution. [Cf.\cite{lan}].
 
Thus let $|x>$ carry the IRR's of $\cb$ on $\ch$ with
$b|x>=b_c(x)|x>, b\in \cb, b_c(x)\in~ \comm$ and
let $I=\int \omega |x><x|$ be the resolution of identity.  Then the
classical
probability density in question is given by $\omega |<x|\psi>|^2$ for a
volume form $\omega$
while the mean value of $b$ in $|\psi>$ is $\int \omega|<x|\psi>|^2
b_c(x)$.

As $|\psi>$ thus is equivalent to a classical probability measure for an
instantaneous measurement (which any way is the only sort of measurement
discussed in usual quantum physics), there is no need to invoke 
``collapse of wave packets'' or similar hypotheses for its interpretation.
The uniqueness of quantum measurement theory then consists in the special
relations it predicts between outcomes of measurements of different
commutative  algebras $\cb_1$ and $\cb_2$.  These relations are often
universal,
being independent of the state vector $|\psi>$.

Such a point of view of quantum physics, or at least a view close to it,
has been advocated especially by Sorkin \cite{sor4}.

Thus we see that instantaneous measurements are linked both to classical
topology and to classical measurement theory.

But surely the notion of {\em instantaneous measurements} can only be an
idealization.  Measurements must be extended in time too, just as they are
extended in space.  But we know of no fully articulated theory of
measurements extended in time, and maintaining quantum coherence during
its
duration, although interesting research about these matters exists \cite{ha1}.

A quantum theory of measurements extended in time, with testable
predictions, could be of fundamental importance. We can anticipate that it
will involve 
noncommutative algebras $\cn$ instead of
commutative algebras, the hermitean form $\psi^\dagger\chi$ for 
the appropriate vectors
$\psi,
\chi$  in the Hilbert space being valued in $\cn$.  
Such quantum theories were
encountered in \cite{bbellst}.  Mathematical tools for their further 
development are
probably available in Noncommutative Geometry \cite{co,col1,con,Landi,ma}.


\begin{thebibliography}{999}

\bibitem{bbellst}  A.P. Balachandran, G. Bimonte, E. Ercolessi, G.
Landi, F. Lizzi, G. Sparano and P. Teotonio-Sobrinho, {\it
Finite Quantum Physics and Noncommutative Geometry} [IC/94/38, DSF-T-2/94
SU-4240-567 (1994), hep-th/9403067], in {\it Proceedings of the XV Autumn
School,`` Particle Physics in the 90's'' }, eds. G. Branco and M. Pimenta,
{\it Nucl. Phys. B} ({\em Proc. Suppl.}) {\bf 37C} (1995) 20; {\it J. Geom.
Phys.} {\bf 18} (1996) 163.


\bibitem{bbms} A.P. Balachandran, G. Bimonte, G. Marmo and A. Simoni, {\it
Nucl.
Phys.} {\bf B446} (1995) 299.


\bibitem{bc}  A.P. Balachandran and L. Chandar, {\it Nucl. Phys.} {\bf
B428}
(1994)435.

\bibitem{rs} M. Reed and B. Simon, {\it Methods of Modern Mathematical
Physics},
Vol. I, {\it Functional Analysis} [Academic Press, 1972].


\bibitem{bdgmss1}  A.P. Balachandran, A. Daughton, Z.-C.
Gu, G. Marmo, R.D. Sorkin,
and A.M. Srivastava, {\it Mod. Phys. Lett.} {\bf A5} (1990) 1575 and {\it
Int. J. Mod. Phys.} {\bf A8} (1993) 2993;  A.P. Balachandran, W.D. McGlinn, L.
O'Raifeartaigh, S. Sen and R.D. Sorkin, {\it Int. J. Mod. Phys.} {\bf A7}
(1992) 6887 [Errata: {\it Int. J. Mod. Phys.} {\bf A9} (1994) 1395]; A.P.
Balachandran, W.D. McGlinn, L. O'Raifeartaigh, S. Sen, R.D. Sorkin and A.M.
Srivastava, {\it Mod. Phys. Lett.} {\bf A7} (1992) 1427.

\bibitem{CoRo} A. Connes and C. Rovelli, {\it Class. Quant. Grav.} {\bf
11} (1994) 2899.

\bibitem{fd} J.M.G. Fell and R.S. Doran, {\it Representations of
*-Algebras,
Locally
Compact Groups and Banach *-Algebraic Bundles} [Academic Press, 1988].

\bibitem{co} A. Connes, {\it Noncommutative Geometry} [Academic Press,
1994].

\bibitem{col1} A. Connes and J. Lott, {\it Nucl. Phys.} {\bf B}
(Suppl.) {\bf 18}
(1990) 29; J.C. V\'{a}rilly and J.M. Gracia-Bond\'{i}a, {\it J. Geom.
Phys.}
{\bf 12} (1993) 223.

\bibitem{con} A. Connes,{\it  Commun. Math. Phys.} {\bf 182}(1996)
155. There are several excellent
accounts
of
Noncommutative Geometry besides refs. 7,8 and this paper. Two such are 
\cite{Landi} and \cite{ma}. They also have references.

\bibitem{Landi} G.Landi, {\it An Introduction to Noncommutative Spaces and
their
Geometry}, hep-th/9701078.

\bibitem{ma} J. Madore, {\it An Introduction to Noncommutative
Differential
Geometry and its Physical Applications}, London Mathematical Society
Lecture Notes Series 206 [Cambridge University Press, 1995]. 

\bibitem{rs2} M. Reed and B. Simon, {\it Methods of Modern Mathematical
Physics},
Vol.
II:{\it Fourier Analysis; Self Adjointness}  
[Academic Press, 1975].

\bibitem{ency} A topological space T is Hausdorff iff there are open sets
$\co(x),
\co(y)$ containing any two points $x,y$ in $T$ such that $\co(x)
\cap \co(y)=\phi$. See {\it Encyclopedic Dictionary of Mathematics},
edited
by Sh\^{o}kichi Iyanaga and Yukiyosi Kawada, translation reviewed by K.O.
May [The M.I.T. Press, 1977].

\bibitem{cor} A. Connes and M.A. Rieffel, {\it Contemporary Mathematics}
{\bf 62}
(1987) 237.


\bibitem{bfss} T. Banks, W. Fischler, S.H. Shenker and L. Susskind,
hep-th/9610043
and
references therein.


\bibitem{fg} J. Fr\"{o}hlich and K. Gawedski, CRM Proceedings and Lecture
Notes,
vol. 7 (1994) 57; J. Fr\"{o}hlich, O. Grandjean and A. Recknagel,
hep-th/9612205.

\bibitem{ma2} D. Marolf, Phys. Lett. {\bf B 392} (1997)287.

\bibitem{fs}  J.L. Friedman and R.D. Sorkin, {\it Phys. Rev. Lett.} {\bf
44}
(1990) 1100; ibid. {\bf 45} (1980) 148;
{\it Gen. Rel. Grav.} {\bf 14} (1982) 615.

\bibitem{sor}  R.D. Sorkin; {\it Introduction to Topological Geons},
in {\it Topological Properties and Global Structure of Space-Time}, eds. P.G.
Bergman and V. de Sabata (Plenum, 1986); {\it Classical Topology
and Quantum
Phases: Quantum Geons}, in  {\it Geometrical and Algebraic
Aspects of Nonlinear
Field Theory}, eds. S. De Filippo, M. Marinaro, G. Marmo and  G. Vilasi
(North-Holland, 1989).

\bibitem{bmss}  A. P. Balachandran, G. Marmo, B. S. Skagerstam and A.
Stern, {\it Classical Topology and Quantum States} [World Scientific, 1991].

\bibitem{ds} F. Dowker and R.D. Sorkin, gr-qc/9609064.

\bibitem{abbjrs} C.Aneziris, A.P.Balachandran,M.Bourdeau, S.Jo,
T.R.Ramadas and 
R.D.Sorkin, {\it Mod.Phys.Lett.} {\bf A4 }(1989)331; {\it Int.J.Mod.Phys.}
{\bf
A4 }
(1989)5459.

\bibitem{see} See especially refs. \cite{sor} and 
\cite{bdgmss1} in this regard.

\bibitem{blms} L. Bombelli, J. Lee, D. Meyer and R.D. Sorkin, {\it Phys.
Rev.
Lett.} {\bf 59} (1987) 521; R.D. Sorkin, {\it Int. J. Theor. Phys.} {\bf
30} (1991) 923. See also
U. Yurtsever, {\it Class.Quant. Grav.} {\bf 11} (1994) 1013.
 

\bibitem{lan} N.P. Landsman, {\it Int. J. Mod. Phys.} {\bf A6} (1991)
5349.

\bibitem{sor4} R.D. Sorkin, {\it Mod. Phys. Letts.} {\bf A9} (1994) 3119.


\bibitem{ha1} J. Hartle, {\it Phys. Rev.} {\bf D51} (1995) 1800;
M.D.
Srinivas, {\it Pram\={a}na} {\bf 47} (1996) 1.

\end{thebibliography}
 \end{document}